\documentclass[12pt]{iopart}
\usepackage{graphicx}
\usepackage{cite}
\usepackage{CJK}
\usepackage[colorlinks,
linkcolor=blue,
anchorcolor=blue,
citecolor=blue,
urlcolor=blue
]{hyperref}

\begin{document}
\begin{CJK*}{UTF8}{gbsn}
\title[]{Effects of resonant magnetic perturbation on locked mode of neo-classical tearing modes}

\author{Weikang TANG (汤炜康), Lai WEI (魏来), Zhengxiong WANG (王正汹), Jialei WANG (王佳磊), Tong LIU (刘桐) and Shu ZHENG (郑殊)}

\address{Key Laboratory of Materials Modification by Laser, Ion, and Electron Beams (Ministry of Education), School of Physics, Dalian University of Technology, Dalian 116024, People’s Republic of China}
\ead{zxwang@dlut.edu.cn}
\vspace{10pt}
% \begin{indented}
% \item[]April 2018
% \end{indented}

\begin{abstract}
Effects of externally applied resonant magnetic perturbation (RMP) on the locked mode of neo-classical tearing mode (NTM) are numerically 
investigated by means of a set of reduced magnetohydrodynamic equations. It is found that for a small bootstrap current fraction, three regimes, 
namely slight suppression regime, small locked island (SLI) regime and big locked island (BLI) regime, are discovered with the increase of RMP strength. 
For a much higher bootstrap current fraction, however, a new oscillation regime appears instead of the SLI regime, although the other regimes still 
remain. The physical process in each regime is analyzed in detail based on the phase difference between NTM and RMP. Moreover, the critical values 
of RMP in both SLI and BLI regimes are obtained, and their dependence on key plasma parameters is discussed as well.
\end{abstract}

\vspace{2pc}
\noindent{\it Keywords}: neo-classical tearing mode, resonant magnetic perturbation, locked mode\\

\noindent(Some figures may appear in colour only in the online journal)
% \submitto{\NF}

% \maketitle

% \ioptwocol

\section{Introduction}\label{sec1}
Neo-classical tearing mode (NTM), as one of the much dangerous macroscopic magnetohydrodynamic (MHD) instabilities, can substantially damage the 
equilibrium magnetic configuration and then contribute to a significant degradation of the plasma confinement\cite{ref1,ref2,ref3}. In general, a ‘seed island’\cite{ref4,ref5} is needed 
to excite the NTM, which can be induced by various kinds of instabilities, such as edge localized mode (ELM)\cite{ref6,ref7,ref8}, tearing mode (TM)\cite{ref9,ref10} and fishbone\cite{ref11,ref12,ref13}. 
The loss of bootstrap current, caused by the flattening of the pressure profile in the inner region of magnetic islands, can further result in a 
destabilization of magnetic islands and even lead to major disruptions\cite{ref14,ref15}. Although achieving a high poloidal beta $\beta_\theta$ \cite{ref16,ref17} is important for improved 
H-mode scenarios in advanced tokamaks, the saturated island width of NTM is proportional to the $\beta_\theta$. Therefore, it is very necessary to control 
the NTM islands in tokamak experiments, because they can seriously restrict the improvement of plasma parameters and limit the performance of tokamak devices. 

Resonant magnetic perturbation (RMP) has different effects on TM or NTM, adjusting its rotation velocity\cite{ref18}, driving magnetic reconnection\cite{ref19} and 
stabilizing the tearing modes\cite{ref20}. Lots of experiments and simulation investigations of RMP effects on TM/NTM were carried out in the previous 
decades\cite{ref21,ref22,ref23}. Yu \emph{et al}. showed that the NTM can be stabilized by an externally applied helical field of a different helicity if the field magnitude 
is sufficiently large\cite{ref24}. Hu \emph{et al}. found that the suppression of the TM by RMP with moderate amplitude is possible and that a small locked 
island (SLI) regime was identified clearly\cite{ref25,ref26}, which was also observed in recent study using two-fluid equations\cite{ref27}. 
However, Hu's work was based on a classical tearing mode model and some neo-classical effects were not considered. Wang \emph{et al}. found that the required RF (radio frequency) current for mode stabilization is reduced 
by about one third if an appropriate RMP is applied\cite{ref28}. Very recently, Choi \emph{et al}. applied a rotating RMP to slow the mode rotation down at a low 
frequency, and then applied the electron cyclotron current drive (ECCD) modulated at the same frequency, so that the avoidance of plasma 
disruption and the re-establishment of the high confined mode were achieved in D\uppercase\expandafter{\romannumeral3}-D\cite{ref29}. 

On the other hand, when the magnitude of a static RMP or residual error field is sufficiently large, the mode frequency of NTM 
can be “arrested” and become the same as the frequency of the RMP (error field), called locked mode (LM)\cite{ref30,ref31,ref32}, which is especially dangerous 
in experiments. These magnetic perturbations can impose electromagnetic torques to brake and ultimately stop the normal rotation of the 
saturated internal magnetic islands that are present in most tokamak plasmas\cite{ref33}. Consequently, the resultant non-rotating magnetic islands 
can rapidly grow to large amplitude and ultimately lead to disruption. Aiming to avoid the LM induced disruption, Fitzpatrick \emph{et al}. 
theoretically and numerically investigated the effect of RMP on scalings of threshold of error field penetration, based on the classical 
TM theory\cite{ref34,ref35,ref36}. Yu \emph{et al} numerically studied the locking of NTM by error fields\cite{ref31}, while the threshold of small locked island was not given.
Although the investigations on the LM of TM were extensively carried out, the detailed simulation study of the nonlinear 
process of LM of NTM is quite limited. Therefore, it is very necessary to have a clear understanding on the effects of RMP on LM of NTM on the basis of 
nonlinear simulations. In particular, the critical value of LM of NTM should be highly valued. 

Motivated by the above reasons, in this work, the effects of RMP on NTM are numerically investigated based on a set of reduced 
magnetohydrodynamic equations. A big locked island (BLI) regime and a small locked island (SLI) regime are discovered and the 
scalings of the critical value of LM of NTM are numerically given. The physical processes of different LM regimes are analyzed 
in detail. The rest of this paper is organized as follows. In section~\ref{sec2}, the modeling equations used in this work are introduced. 
In section~\ref{sec3}, numerical results and physical discussions are presented. Finally, the paper is summarized and conclusions are drawn in section~\ref{sec4}.

\section{Modeling equations}\label{sec2}
The nonlinear evolution of NTM with the existence of RMP, in this paper, is studied by means of a set of reduced MHD equations\cite{ref37,ref38,ref39,ref40,ref41,ref42} 
in the cylinderical geometry ($r$, $\theta$, $z$). The normalized equations, containing the evolution of the vorticity, magnetic flux, and plasma pressure, 
are as follows
\begin{equation}
\frac{\partial u}{\partial t} = [u,\phi] + [j,\psi] + \partial_zj + R^{-1}\nabla^2_\perp u,\label{eq1}
\end{equation}
\begin{equation}
\frac{\partial \psi}{\partial t} = [\psi,\phi] - \partial_z\phi - S^{-1}_{\rm A}(j - j_{\rm b}) + E_{z0},\label{eq2}
\end{equation}
\begin{equation}
\frac{\partial p}{\partial t} = [p,\phi] + \chi_\parallel\nabla^2_\parallel p + \chi_\perp\nabla^2_\perp p + S_0,\label{eq3}
\end{equation}
where $\phi$ and $\psi$, respectively, are the stream function and magnetic flux. Vorticity and plasma current density can be expressed as the form of 
$u=\nabla^2_\perp \phi$ and $j=-\nabla^2_\perp \psi$, respectively. Here, $p$ represents the plasma pressure. $j_{\rm b}=-f_{\rm b}\frac{\sqrt\varepsilon}{B_\theta}\frac{\partial p}{\partial r}$
is the bootstrap current with a fraction $f_{\rm b}(r,\beta) = \int_0^a j_{\rm b0}r{\rm d}r/\int_0^a j_{z0}r{\rm d}r$, which accounts for the ratio of the bootstrap current to total current 
along the axis. Normalized by $a^2/\tau_{\rm a}$, $\chi_\parallel$ and $\chi_\perp$ are the parallel and perpendicular transport coefficients, respectively. 
The radial coordinate $r$, time $t$ and velocity $v$ are normalized by the plasma minor radius $a$, Alfv\'{e}n time $\tau_{\rm a}=\sqrt{\mu_0\rho}a/B_0$ 
and Alfv\'{e}n speed $v_{\rm a}=B_0/\sqrt{\mu_0\rho}$, respectively.  
$R=\tau_\nu/\tau_{\rm a}$ is the Reynolds number and $S_{\rm A}=\tau_\eta/\tau_{\rm a}$ represents the magnetic Reynolds number in equations (\ref{eq1}) and (\ref{eq2}), where 
$\tau_\nu=a^2/\nu_{\rm c}$ and $\tau_\eta=a^2\mu_0/\eta_{\rm c}$ are viscosity diffusion time and resistive diffusion time, respectively. 
In order to restore the dissipation of the initial profiles of Ohm current $j_{\rm Ohm}(r)$ and pressure $p_0(r)$, source terms $E_{z0}=S^{-1}_{\rm A}j_{\rm Ohm}=S^{-1}_{\rm A}(j_{z0}-j_{\rm b0})$ 
and $S_0=-\chi_\perp\nabla^2_\perp p_0$ are utilized, where $j_{\rm b0}$ and $j_{z0}$ are the $z$-component of the initial bootstrap and total current density, respectively. 
In cylinderical geometry the Poisson brackets in equations~(\ref{eq1})$-$(\ref{eq3}) can be written as
\begin{equation} 
[f,g]=\nabla f\times\nabla g\cdot \hat{z}=\frac{1}{r}(\frac{\partial f}{\partial r}\frac{\partial g}{\partial \theta}-\frac{\partial g}{\partial r}
\frac{\partial f}{\partial \theta}).\label{eq4}
\end{equation}
Every single variable $f(r,\theta,z,t)$ in equations~(\ref{eq1})$-$(\ref{eq3}) can be splited into two components as $f = f_0(r) + \widetilde{f}(r,\theta,z,t)$, 
where $f_0$ is the time-independent equilibrium profile and $\widetilde{f}$ is the time-dependent perturbed part. 
Owning to the periodic symmetry in the poloidal and toroidal directions, the perturbed fields can be Fourier-transformed in the form of
\begin{equation}
\widetilde{f}(r,\theta,z,t)=\frac{1}{2}\sum_{m,n}\widetilde{f}_{m,n}(r,t){\rm e}^{({\rm i}m\theta-{\rm i}nz/R_0)},\label{eq5}
\end{equation}
For specific profiles of $\phi_0, \psi_0$, and $p_0$, equations~(\ref{eq1})$-$(\ref{eq3}) can be solved by an initial value
code: MD code (MHD@Dalian Code)\cite{ref43,ref44,ref45}. The finite difference method is used in the radial direction and the pseudo-spectral 
method is employed for the poloidal and the toroidal directions ($\theta,\zeta=-z/R_0$). 
For a better acuracy and stability, a semi-implicit scheme, two-step predictor-corrector method, is applied in the time advancement. 
The MD code has been benchmarked with the codes adopted in reference \cite{ref46} in linear and nonlinear calculations with a variety of configurations. 

\begin{figure}
\centering
\includegraphics[width=1.\textwidth]{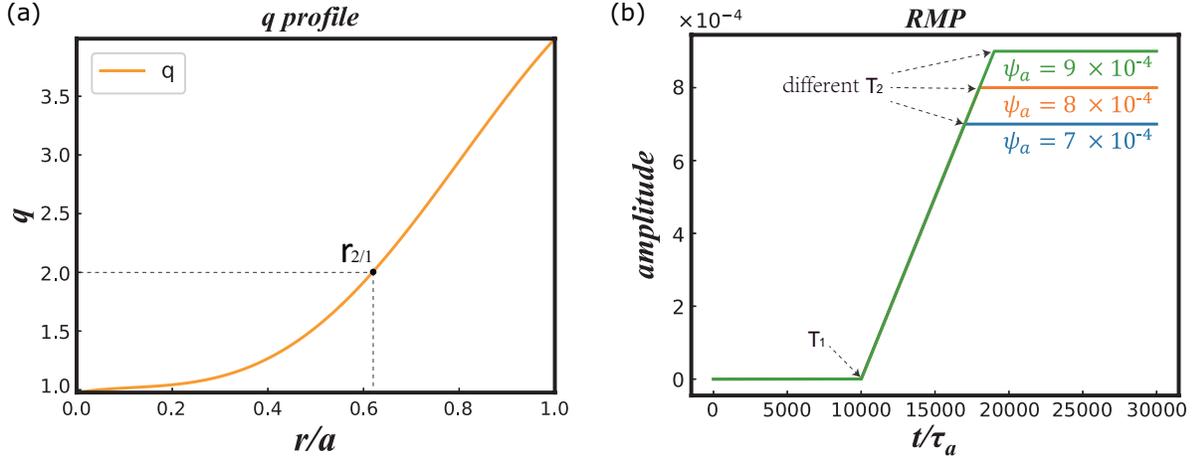}
\caption{(a) Safety factor $q$ profile and (b) RMP configuration applied in this work. As (b) illustrated, amplitude of RMP is modulated by changing the flattop value of RMP.}
\label{fig1}
\end{figure}

In this work, the plasma rotation is considered by setting $v_0(r)=\frac{\rm d}{{\rm d}r}\phi_0(r)=\Omega_0(1-r)$, where $\Omega_0$ is the plasma rotation frequency on the 
magnetic axis. The initial profiles of safety factor $q$ is shown in figure \ref{fig1}(a), in which the $m/n$ = 2/1 tearing mode without bootstrap current 
is stable ($\Delta'<0$). The effect of RMP with $m/n$ is taken into account by the boundary condition
\begin{equation}
\widetilde{\psi}_{m,n}(r = 1) = \psi_a(t){\rm e}^{({\rm i}m\theta-{\rm i}nz/R_0)}.\label{eq6}
\end{equation}
The time-dependent RMP $\psi_a(t)$ in equation (\ref{eq6}) is set as 
\begin{equation}
\psi_a(t) = \cases{0&$0<t\le T_{1}$\\
\frac{{\rm d}\psi}{{\rm d}t}\cdot(t-T_{1})&$T_{1}<t\le T_2$\\
\psi_{\rm RMP}&$t>T_2$\\},\label{eq7}
\end{equation}
where $T_1$ is the time when turning on the RMP, and ${\rm d}\psi/{\rm d}t$ is the growth rate of the RMP. In our simulation, we keep the growth rate of the RMP as a 
constant and modulate the amplitude of RMP by changing the flattop value of RMP as figure \ref{fig1}(b) illustrated. It should be pointed out that, in 
a real tokamak, the toroidal rotation is prevailing and much stronger than the poloidal one, whereas only the poloidal rotation is considered in this 
work due to the large aspect ratio approximation. The electromagnetic force exerted in the poloidal direction is $(n/m)(r_s/R)$ times larger than that in 
toroidal direction, and the speed in toroidal direction should be $(n/m)(r_s/R)$ times larger than the poloidal one for having an equivalent rotation frequency. 
Therefore, the locking threshold in the toroidal direction can be estimated by multiplying such a factor $[(n/m)(r_s/R)]^2$.
   
\begin{figure}
\centering
\includegraphics[width=1.\textwidth]{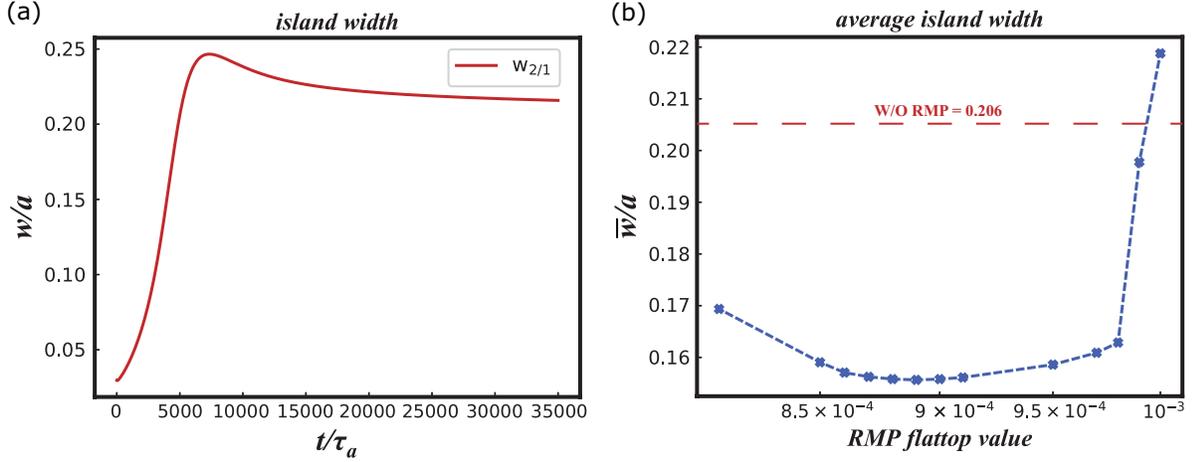}
\caption{(a) Nonlinear evolution of the 2/1 magnetic island width without RMP. (b) Average island widths with different RMP amplitudes. The 
common parameters are set as $f_{\rm b}=0.1$, $R^{-1}=1\times10^{-4}$, $S_{\rm A}^{-1}=5\times10^{-5}$, $\chi_\parallel=10$ and $\chi_\perp=1\times10^{-6}$.}
\label{fig2}
\end{figure}
\section{Numerical results}\label{sec3}
\subsection{Effects of RMP amplitude}
\begin{figure}
\centering
\includegraphics[width=.6\textwidth]{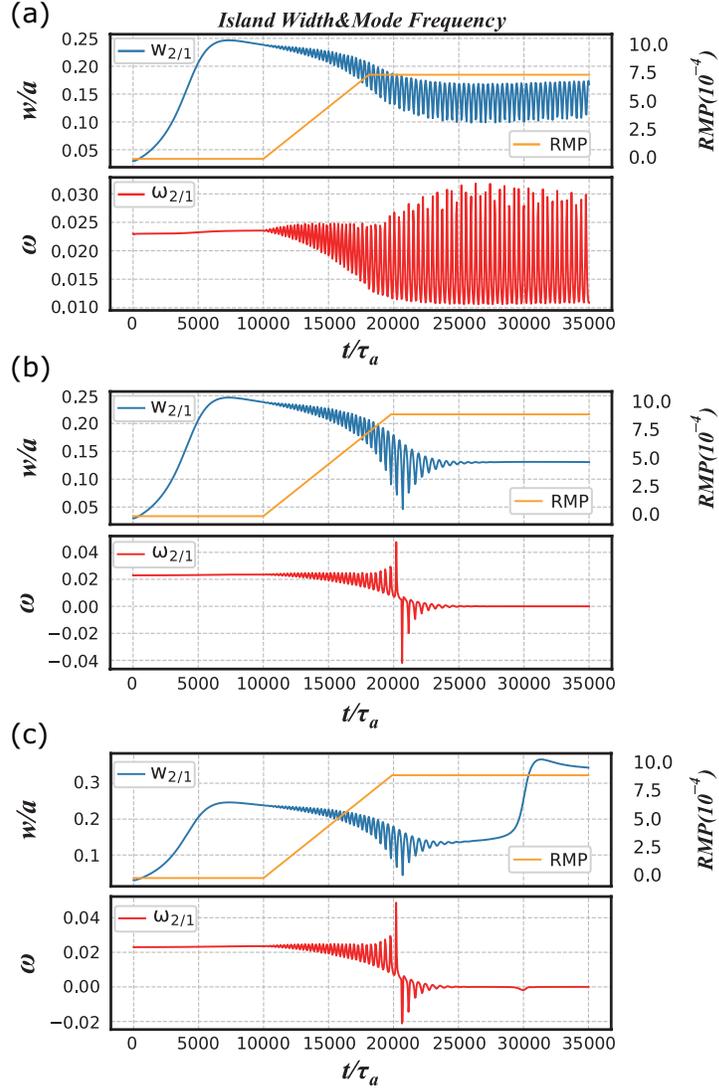}
\caption{The three typical processes for a small bootstrap current fraction $f_{\rm b}=0.10$: (a) slight suppression, (b) small locked island, (c) big locked island. The corresponding RMP amplitudes are 8.1$\times10^{-4}$, 9.8$\times10^{-4}$ and 9.9$\times10^{-4}$, respectively.}
\label{fig3}
\end{figure}
\begin{figure}
\centering
\includegraphics[width=.6\textwidth]{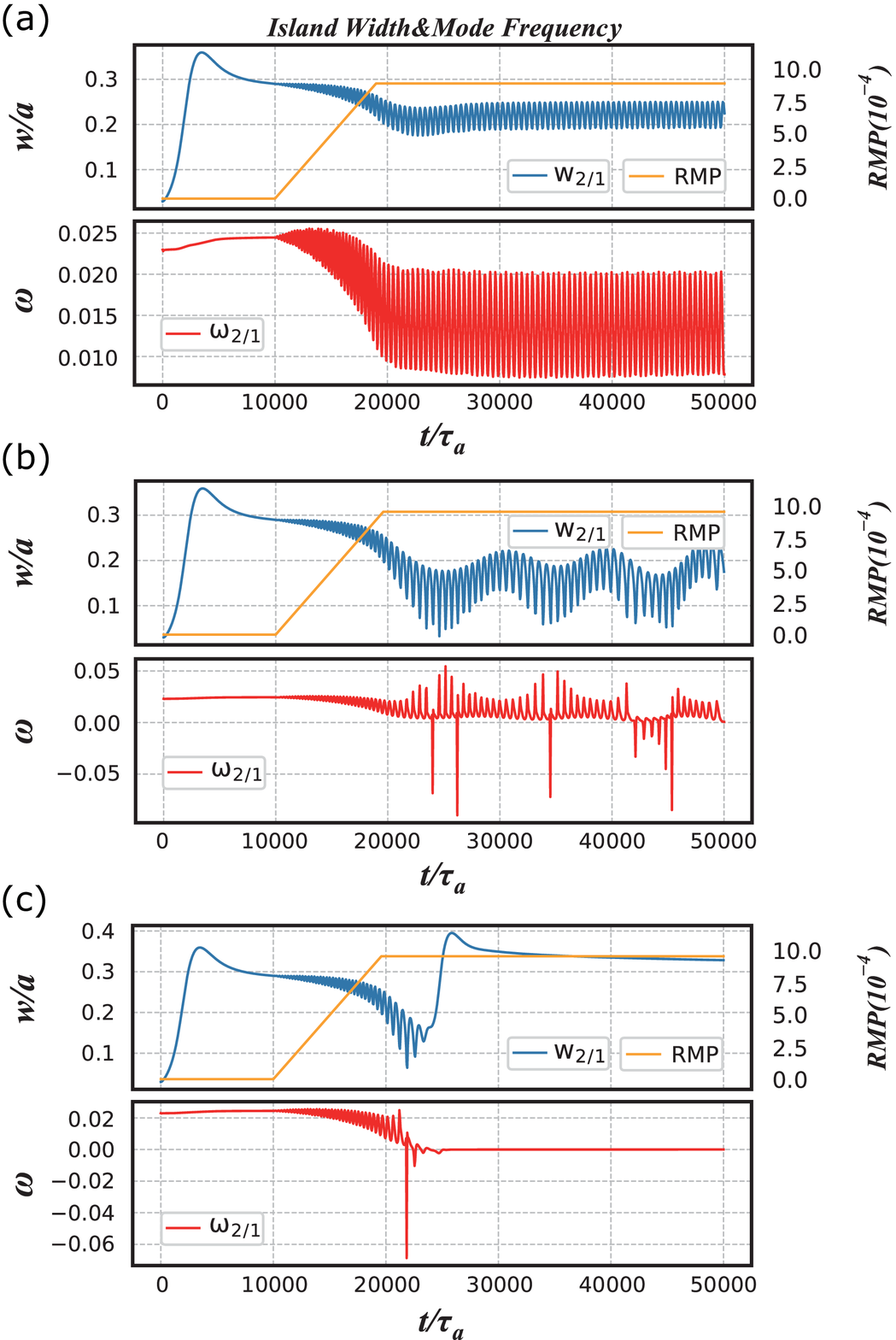}
\caption{The three typical processes for a large bootstrap current fraction $f_{\rm b}=0.25$: (a) slight suppression, (b) oscillating phase, (c) big locked island. The corresponding RMP amplitudes are 9.0$\times10^{-4}$, 9.6$\times10^{-4}$ and 9.8$\times10^{-4}$, respectively.}
\label{fig4}
\end{figure}    
\begin{figure}
\centering
\includegraphics[width=.6\textwidth]{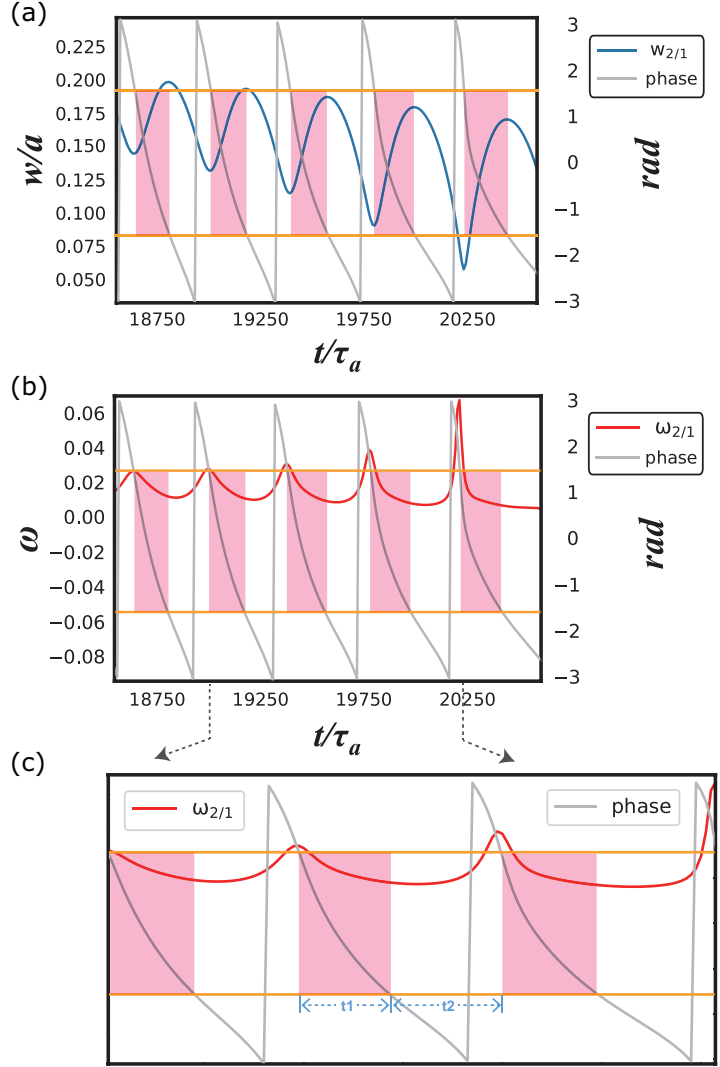}
\caption{(a) Magnetic island widths together with the phase difference and (b) mode frequency together with the phase difference in the early suppression period (c) magnification of three phase periods in (b). The yellow horizon lines are $-\pi/2$ and $\pi/2$ respectively. The input parameters are set as $f_{\rm b}=0.10$, $\psi_{\rm RMP}=9.8\times 10^{-4}$.}
\label{fig5}
\end{figure}
\begin{figure}
\centering
\includegraphics[width=1.\textwidth]{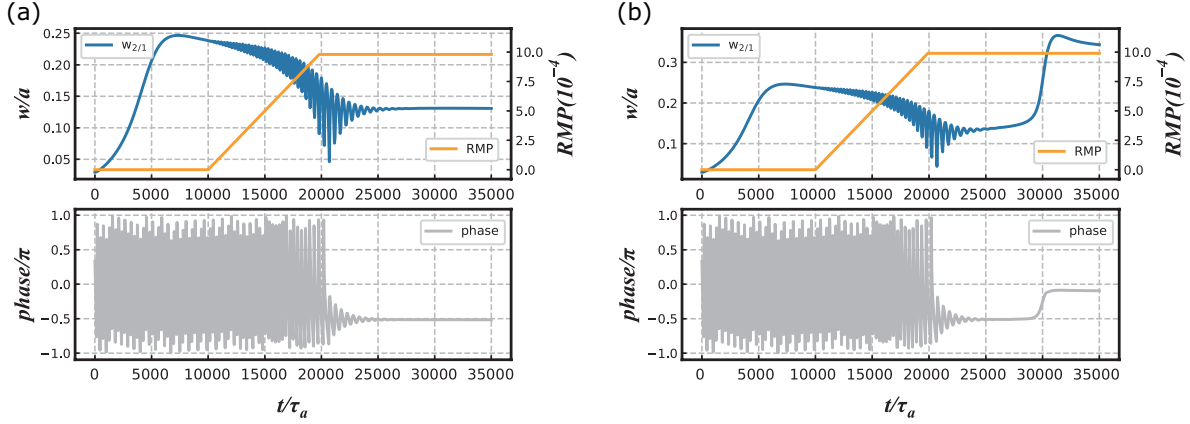}
\caption{The phase analysis of the small locked island case (a) and big locked island case (b) in figure \ref{fig3}. These two kinds of locked modes are locked to the different phases. Note that there is a jump of the phase when big locked island happens.}
\label{fig6}
\end{figure}
\begin{figure}
\centering
\includegraphics[width=.95\textwidth]{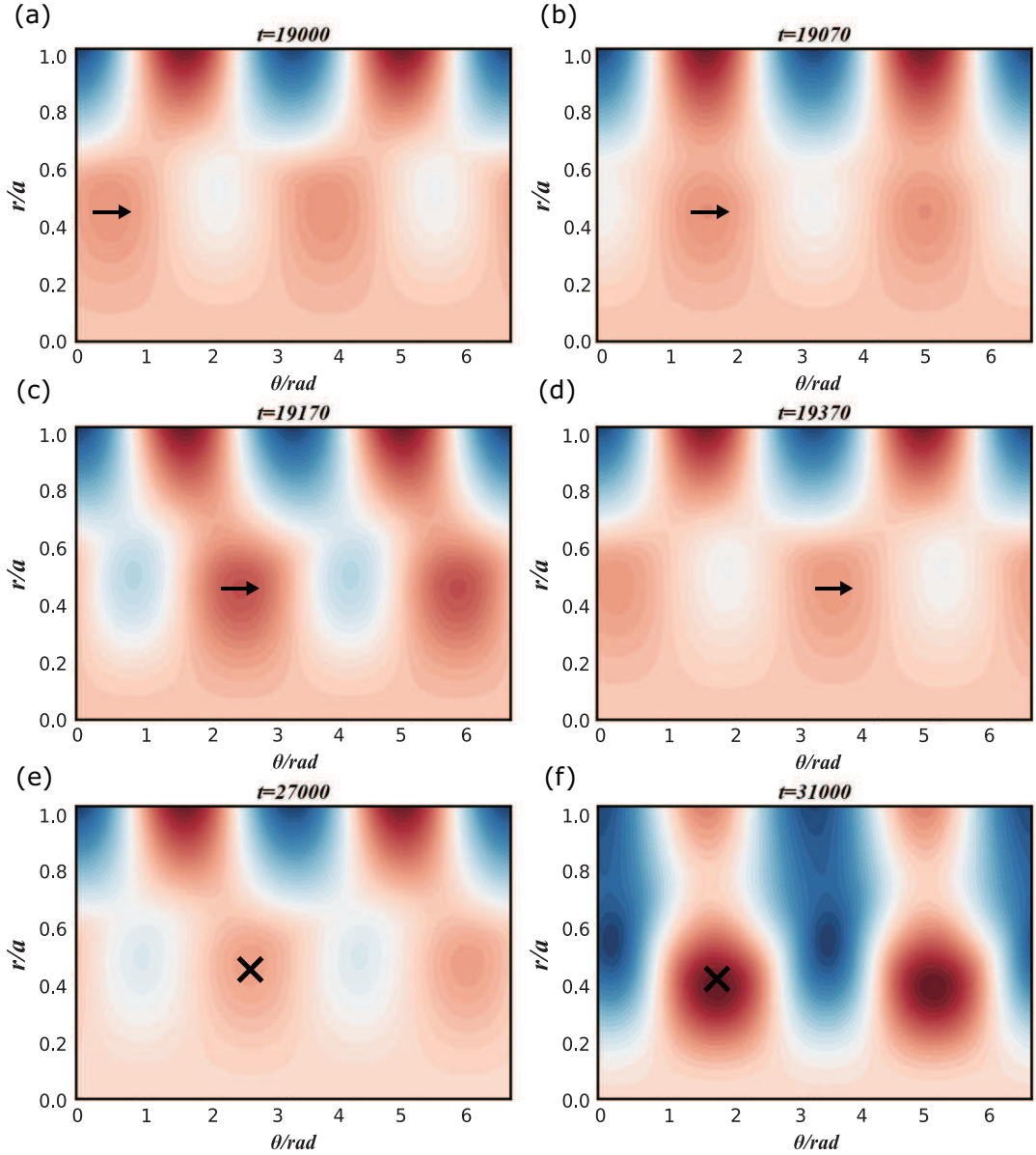}
\caption{Perturbed magnetic flux in the process of figure \ref{fig6}(b). The four pictures with arrows are the perturbed magnetic flux in an entire phase period before mode locked and the arrows indicate the direction where it moves. The two pictures with crosses show the perturbed magnetic flux after locked mode, which means the mode is static.} 
\label{fig7}
\end{figure}
\begin{figure}
\centering
\includegraphics[width=.5\textwidth]{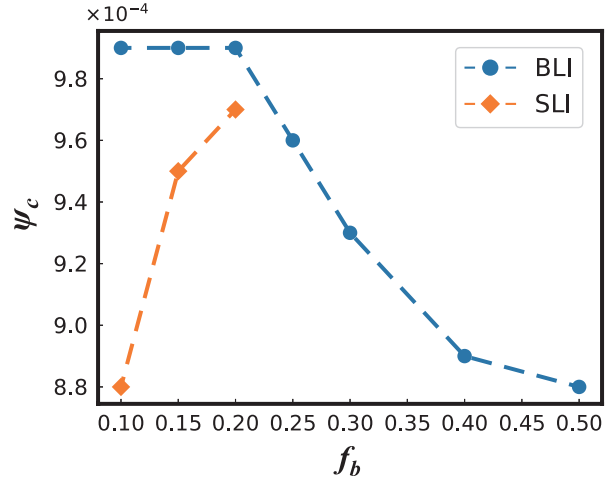}
\caption{The critical value of locked mode $\psi_{\rm c}$ versus the bootstrap current fraction $f_{\rm b}$. All scans are performed with $R^{-1}=1\times10^{-4}$, $S_{\rm A}^{-1}=5\times10^{-5}$, $\chi_\parallel=10$, $\chi_\perp=1\times10^{-6}$.}
\label{fig8}
\end{figure}
\begin{figure}
\centering
\includegraphics[width=.5\textwidth]{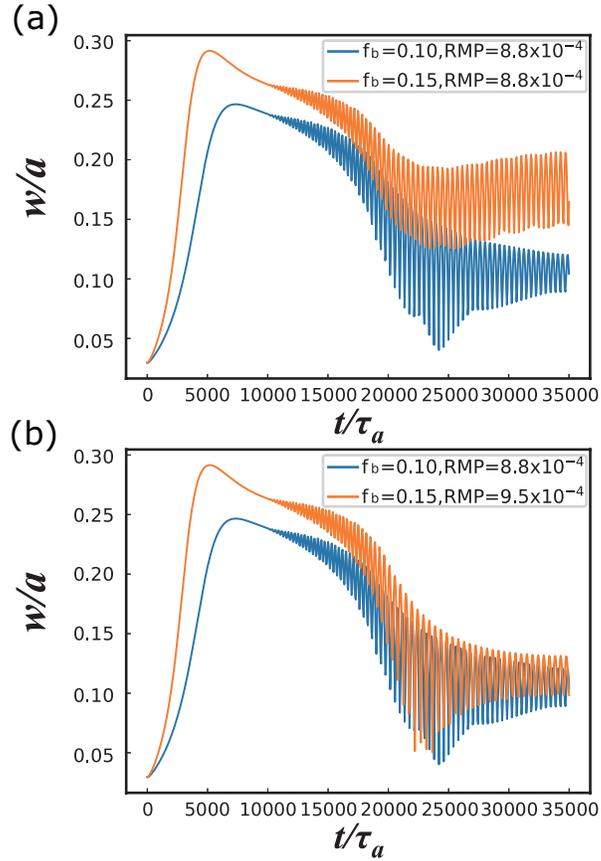}
\caption{Comparison of the nonlinear evolution of island widths for different $f_{\rm b}$. $8.8\times 10^{-4}$ and $9.5\times 10^{-4}$ are the critical RMP values of SLI for $f_{\rm b}=0.10$ and $f_{\rm b}=0.15$ respectively as shown in figure \ref{fig8}.} 
\label{fig9}
\end{figure}
\begin{figure}
\centering
\includegraphics[width=.5\textwidth]{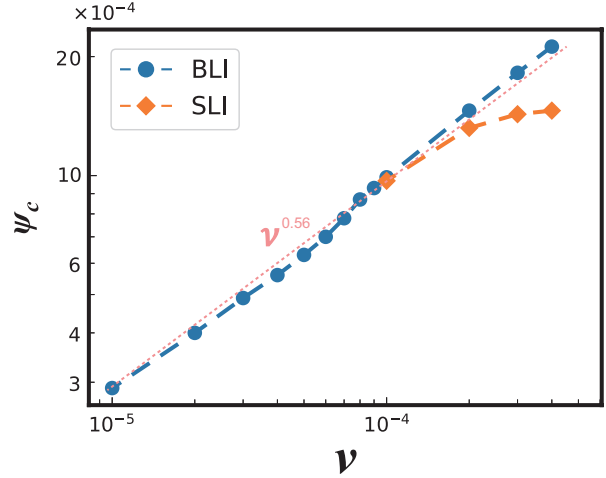}
\caption{Dependence of the plasma viscosity $\nu$ and the critical value of locked mode $\psi_{\rm c}$. All scans are carried out with $f_{\rm b}=0.2$, $S_{\rm A}^{-1}=5\times10^{-5}$, $\chi_\parallel=10$, $\chi_\perp=1\times10^{-6}$. }
\label{fig10}
\end{figure}
\begin{figure}
\centering
\includegraphics[width=.5\textwidth]{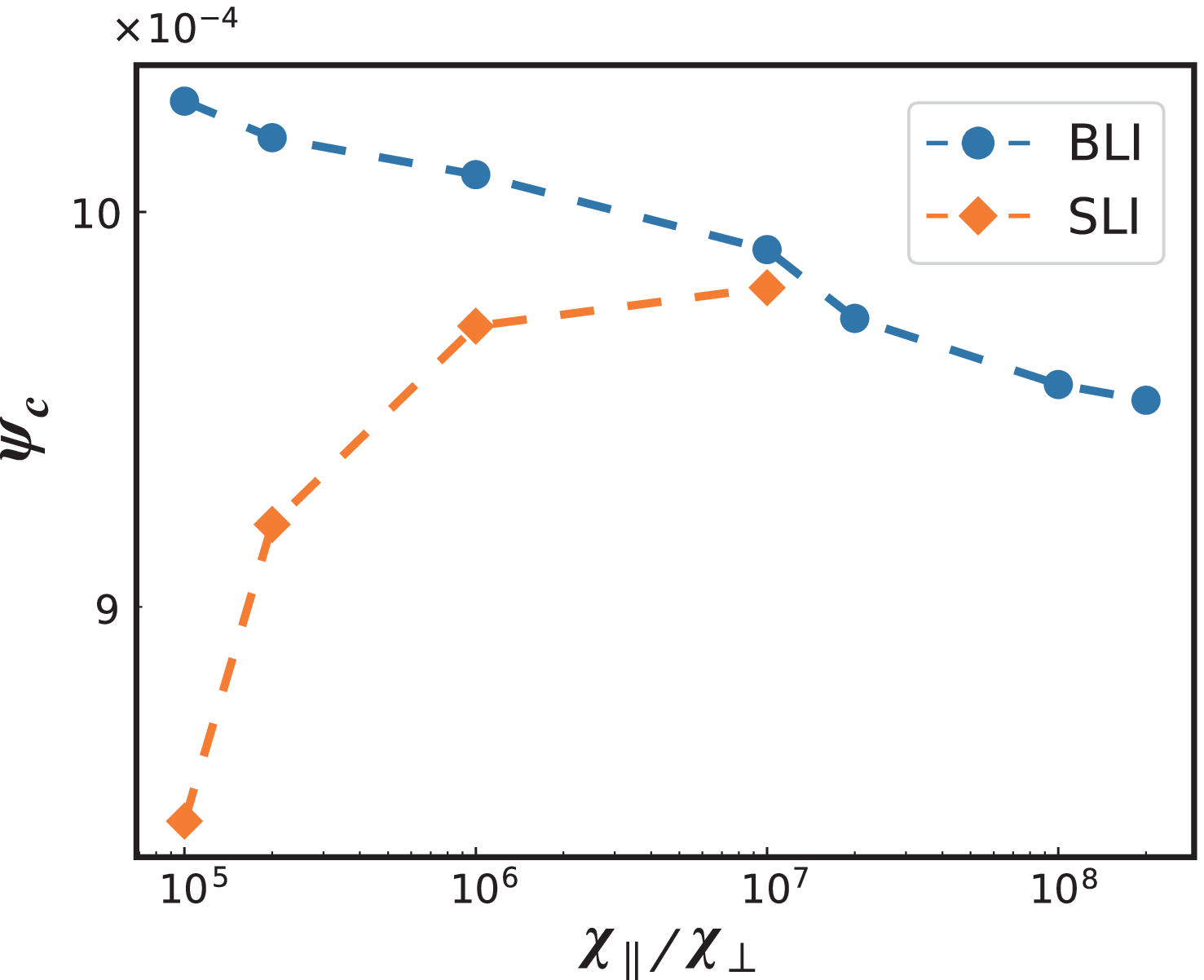}
\caption{The critical value of locked mode $\psi_{\rm c}$ versus the ratio of the parallel transport coefficient to perpendicular transport coefficient ${\chi_\parallel/\chi_\perp}$. All scans are conducted with $f_{\rm b}=0.2$, $R^{-1}=1\times10^{-4}$, $S_{\rm A}^{-1}=5\times10^{-5}$.}
\label{fig11}
\end{figure}

For a comprehensive understanding of the effects of RMP on NTMs, the effects of different RMP amplitudes on NTMs are investigated in this section. The 
common parameters are set as $f_{\rm b}=0.1$, $R^{-1}=1\times10^{-4}$, $S_{\rm A}^{-1}=5\times10^{-5}$, $\chi_\parallel=10$ and $\chi_\perp=1\times10^{-6}$. 
The nonlinear evolution of NTM without RMP is given in figure \ref{fig2} as a baseline case. It is seen in figure \ref{fig2}(a) that the $m/n=2/1$ magnetic island 
saturates at around $t = 20000$ and the saturate island width is about 0.2. During the whole evolution, the average island width $\bar{w}$ is 0.206.

Firstly, a small bootstrap current fraction $f_{\rm b}=0.1$ is taken into account. Figure \ref{fig2}(b) gives the average island widths in the presence of different 
RMP amplitudes. It is found that with the increase of RMP value, $\bar{w}$ first decreases, indicating a stabilizing effect of RMP. However, once $\psi_{\rm RMP}$ exceeds 
a threshold, $\bar{w}$ immediately increases again. The processes associated with the relation between $\bar{w}$ and  $\psi_{\rm RMP}$ can be divided into three 
regimes, which are illustrated in figure \ref{fig3} and discussed in detail as follows.
\begin{enumerate}
\item Slight suppression regime: when the RMP is turned on at $t =10000$, the island width and the mode frequency start to oscillate due to the electromagnetic 
torque applied by the boundary magnetic perturbation. As a result, the NTM can be slightly stabilized for a low $\psi_{\rm RMP}$, such as  $\psi_{\rm RMP}=8.1 \times 10^{-4}$. 
\item Small locked island regime: for a moderate RMP, the NTM can be further stabilized and the island width is reduced to 0.13. The mode frequency drops slowly 
and then is suddenly locked to the RMP when the plasma rotation has been reduced to one half of its original value. 
\item Big locked island regime: for a sufficiently large RMP, the field penetration occurs accompanied with an explosive growth of island width, and 
the mode frequency is locked to the static magnetic field, which is known as the so-called locked mode.
\end{enumerate}

Secondly for a higher bootstrap current fraction, however, the regimes determined by different RMP amplitudes are different. As shown in 
figure \ref{fig4}, there are also three typical regimes for $f_{\rm b}=0.25$: (i) slight suppression regime; (ii) oscillation regime; (iii) big 
locked island regime. Although the (i) and (iii) regimes are almost the same as those of $f_{\rm b}=0.1$, a new oscillation regime (ii) appears 
instead of the small locked island regime (ii) of $f_{\rm b}=0.1$.

To understand the effects of RMP on NTM deeply, phase analysis has been carried out. Figure \ref{fig5} gives the phase of the NTM, the 
island width and the mode frequency in the early suppression period, which can be observed in all regimes mentioned above. The phase 
of NTM is equal to the phase difference between the NTM and the RMP, since the phase of the static RMP is zero. In figures \ref{fig5}(a)$-$(c), 
the gray line is the phase of NTM, the blue line is the island width, and the red line is the mode frequency. The yellow horizon lines 
are $\pi/2$ and $-\pi/2$ respectively. When the phase difference is at $\Delta\Phi<|\pi/2|$ (pink area), the RMP has a destabilizing 
effect on NTM. Oppositely, for $\Delta\Phi>|\pi/2|$ (white area), the RMP has a stabilizing effect. Moreover, the mode frequency of 
the NTM also changes periodically. The RMP has a deceleration effect on the mode frequency when $\Delta\Phi<|\pi/2|$ (pink area). 
For $\Delta\Phi>|\pi/2|$ (white area), the RMP has an acceleration effect. It can be clearly noted in figure \ref{fig5}(c) that time 
interval $t_2$ in the white area is longer than $t_1$ in the pink area, leading to a net stabilizing effect on the island in an entire 
period. On the other hand, $t_1 < t_2$ also indicates that the average mode frequency in the pink area is larger than that in the white 
area. As the amplitude of RMP gets higher, the electromagnetic torque becomes much bigger, which can be judged from the slopes of 
the mode frequency growth shown in figure \ref{fig5}. Once the electromagnetic force exceeds a threshold, the direction of the rotation 
can reverse. Then the mode can no longer travel through the whole phase region and the angular velocity is slowly damped, hence the 
SLI in the (ii) regime finally occurs, as shown in figure \ref{fig6}(a).

It should be noted in figures \ref{fig6}(a) and (b) that the phase of SLI is locked to $-$0.52$\pi$, which is essentially different from the 
BLI regime where the phase is locked to $-$0.10$\pi$ (almost the phase of the static RMP). For the SLI, the phase difference is in 
$\Delta\Phi>|\pi/2|$, where the RMP has a stabilizing effect on NTM. But for the BLI, the phase difference is in $\Delta\Phi<|\pi/2|$, 
where the RMP has a destabilizing effect on NTM. That is the reason why there are two kinds of locked mode with significantly different 
island widths for SLI and BLI. 

Figure \ref{fig7} shows the evolution of the perturbed magnetic flux of BLI in figure \ref{fig6}(b). Figures \ref{fig7}(a)$-$(d) give continuous 
motion slices of perturbed magnetic flux in one entire phase period. It can be found that when the NTM and the RMP are in the same phase, 
the RMP can enhance the magnetic reconnection, as shown in figures \ref{fig7}(b) and (c). But when they are in anti-phase, the RMP can 
prevent the magnetic reconnection, as shown in figures \ref{fig7}(a) and (d). Hence the magnetic island width periodically changes while 
the mode is rotating. Then the SLI occurs at about $t=27000$, and the island width is still small, as shown in figure \ref{fig7}(e), in which 
the NTM is static in the anti-phase of the RMP. But during $t=29000-31000$, as shown in figure \ref{fig7}(f), the NTM is dragged to the same 
phase of the RMP by the electromagnetic torque, which leads to an explosive growth of the island width due to the continuous strong 
magnetic reconnection by the RMP.

\subsection{The dependence of critical value of LM and plasma parameters}
In this section, we analyze the critical RMP values in both SLI and BLI cases. Figure \ref{fig8} shows the critical value $\psi_{\rm c}$ versus $f_{\rm b}$, 
where the blue and yellow lines represent the critical values for BLI and SLI, respectively. 

The $\psi_{\rm c}$ of BLI decreases with the increase of $f_{\rm b}$ due to the destabilizing effect of the bootstrap current. Since for a larger $f_{\rm b}$, the NTM 
becomes more unstable and the island width becomes larger, so that a lower RMP value is needed to lock the NTM. Interestingly, it is noticed 
that the $\psi_{\rm c}$ of SLI increases with the increase of $f_{\rm b}$.

To make clear the $\psi_{\rm c}$ dependence difference, the nonlinear evolution of magnetic island widths are compared for different $f_{\rm b}$ values. 
In figure \ref{fig9}(a), it can be seen that for the same RMP amplitude of $\psi_{\rm RMP}=8.8\times 10^{-4}$, the case of $f_{\rm b}=0.15$ is still in 
the slight suppression regime, while the case of $f_{\rm b}=0.10$ enters the SLI regime. Here for $f_{\rm b}=0.10$, $\psi_{\rm RMP}=8.8\times 10^{-4}$ is the 
critical value of SLI, as shown in figure \ref{fig8}. Figure \ref{fig9}(b) gives the evolution of the island widths for $f_{\rm b}= 0.10$, 
$\psi_{\rm RMP}=8.8\times 10^{-4}$ and $f_{\rm b}= 0.15$, $\psi_{\rm RMP}=9.5\times 10^{-4}$. For $f_{\rm b} =0.15$, $\psi_{\rm RMP}=9.5\times 10^{-4}$ is the critical 
value of SLI, as shown in figure \ref{fig8}.

It is found that when SLI occurs, the island width of $f_{\rm b}=0.15$ is suppressed to almost the same level as that of the $f_{\rm b}=0.10$. In our 
simulations, the island widths of SLI are found to be always around 0.1. It is revealed that for triggering a SLI, the island width should 
be small enough, indicating the existence of the critical value of the NTM island width. The correspondence existence of SLI in the classical 
TM has been identified in simulations and J-TEXT experiments\cite{ref25}. Because a larger RMP value is required to suppress the island width to a small 
level, the critical value of SLI increases with the increase of $f_{\rm b}$. For an even larger $f_{\rm b}$, however, the BLI has occurred before the island 
width is suppressed to the level for triggering the SLI. Finally, the SLI disappears with the increase of $f_{\rm b}$, and thus only BLI remains.

Dependence of critical value of locked mode on plasma viscosity is discussed, since the plasma viscosity is a key parameter affecting the 
mode locking. Figure \ref{fig10} gives $\psi_{\rm c}$ versus plasma viscosity $\nu$. It is seen that for both BLI and SLI, $\psi_{\rm c}$ increases with 
the increase of $\nu$. It is reasonable that the large viscosity denotes a large viscous torque, which can balance the electromagnetic torque. 
Therefore, a larger electromagnetic force generated by RMP is needed to lock the NTM in the large viscosity region. Besides, it is noted 
that the SLI appears only in the large viscosity region. On the other hand, the scaling of the BLI with viscosity is numerically obtained 
as $\psi_{\rm crit} \sim \nu^{0.56}$, which is very close to the theoretical one $\psi_{\rm crit} \sim \nu^{7/12} $\cite{ref35}. 

Figure \ref{fig11} illustrates $\psi_{\rm c}$ versus the ratio of parallel to perpendicular transport coefficient ${\chi_\parallel/\chi_\perp}$, 
since ${\chi_\parallel/\chi_\perp}$ is very important for determining the saturated width of NTM island. Because the increase of 
${\chi_\parallel/\chi_\perp}$ normally increases the saturated island width, $\psi_{\rm c}$ of BLI decreases. On the other hand, the SLI can 
only appear in the small ${\chi_\parallel/\chi_\perp}$ region. In fact, the effect of ${\chi_\parallel/\chi_\perp}$ on $\psi_{\rm c}$ 
is similar to that of bootstrap current fraction $f_{\rm b}$ shown in figure \ref{fig8}. Finally, it is indicated in figures \ref{fig8},\ref{fig10} and \ref{fig11} 
that the SLI tends to occur, when the NTM is in less unstable regimes.        

\section{Summary}\label{sec4}
In summary, the effects of externally applied static RMP on LM of NTM are numerically investigated by means of a set of reduced 
magnetohydrodynamic equations. A small locked island (SLI) regime and a big locked island (BLI) regime are discovered. Effects 
of some significant plasma parameters on the properties of the SLI and BLI are analyzed in detail. The main results can be summarized as follows.
\begin{enumerate}
\item For the NTM with a small bootstrap current fraction $f_{\rm b}$, three regimes, namely slight suppression regime, SLI regime 
and BLI regime, are found with the increase of RMP amplitude. In the case with a low RMP amplitude, the NTM can be always 
stabilized. With the increase of RMP strength, two kinds of locked mode with qualitatively different properties are observed. 
In the SLI/BLI regime, the stabilized/destabilized islands of NTMs are locked to the RMP with phase difference $\Delta \Phi$ being greater/less than $\pi/2$.
\item For the NTM with a large bootstrap current fraction $f_{\rm b}$, islands of NTM can be slightly suppressed with a low RMP 
amplitude, which is the same as the result in the small $f_{\rm b}$ case. However, for a moderate RMP amplitude, an oscillating 
regime appears instead of the SLI regime. For a sufficiently large RMP amplitude, the BLI occurs. Thus, there are also 
three regimes, namely sight suppression regime, oscillation regime and BLI regime.
\item Dependence of the critical value of RMP $\psi_{\rm c}$ for BLI and SLI on the bootstrap current fraction $f_{\rm b}$, plasma 
viscosity $\nu$, and the ratio of parallel to perpendicular transport coefficient ${\chi_\parallel/\chi_\perp}$ are 
further discussed. It is found that the $\psi_{\rm c}$ of BLI decreases as $f_{\rm b}$ increases, since the electromagnetic force 
is proportional to the amplitude of the NTM which increases with the increase of $f_{\rm b}$. However, the $\psi_{\rm c}$ of SLI 
unexpectedly increases with the increase of $f_{\rm b}$. This is due to the fact that, in this work, there is a critical 
value of island width for SLI, a higher RMP strength is needed to suppress the NTM to a necessarily small amplitude 
to trigger the SLI. For plasma viscosity $\nu$, $\psi_{\rm c}$ of both SLI and BLI increase with the increase of $\nu$, 
since the viscous torque, to a great extent, can balance the electromagnetic torque. And only in the large $\nu$ region, 
the SLI occurs. Finally, like the effects of $f_{\rm b}$, the $\psi_{\rm c}$ of BLI/SLI decreases/increases with the increase 
of ${\chi_\parallel/\chi_\perp}$. The SLI can only be found in the small ${\chi_\parallel/\chi_\perp}$ region. 
It is found through the above results that SLI tends to occur when the NTM is in less unstable regimes.
\end{enumerate}

\section*{Acknowledgements}
The author W. Tang is indebted to Dr. Q. Yu for helpful discussions. Useful code benchmark work with T. Hender is greatly appreciated. 
The authors also acknowledge the Supercomputer Center of Dalian University of Technology for providing computing resources. This work 
was supported by the National Key R\&D Program of China (Nos. 2017YFE0301900 and 2017YFE0301100), National Natural Science Foundation of China 
(No. 11675083), the Fundamental Research Funds for the Central Universities (Nos. DUT18ZD101 and DUT17LK38), and the Dalian Youth Science 
and Technology Project Support Program (No. 2015R01).

\end{CJK*}

\section*{References}
\begin{thebibliography}{10}
\bibitem{ref1} ITER 1999 \textit{Nucl. Fusion} \href{http://iopscience.iop.org/article/10.1088/0029-5515/39/12/301/meta} {\textbf{39} 2137}
\bibitem{ref2} Hender T C \emph{et al} 2007 \textit{Nucl. Fusion} \href{http://stacks.iop.org/0029-5515/47/i=6/a=S03?key=crossref.0b9322e42a59b955269531e35f39b3bb} {\textbf{47} S128} 
\bibitem{ref3} Maraschek M 2012 \textit{Nucl. Fusion} \href{http://iopscience.iop.org/article/10.1088/0029-5515/52/7/074007} {\textbf{52} 074007}
\bibitem{ref4} Gorelenkov N N \emph{et al} 1996 \textit{Phys. Plasmas} \href{https://aip.scitation.org/doi/abs/10.1063/1.871614} {\textbf{3} 3379}
\bibitem{ref5} Wang W \emph{et al} 2018 \textit{Plasma Sci. Technol.} \href{https://doi.org/10.1088/2058-6272/aab48f} {\textbf{20} 075101}
\bibitem{ref6} Zohm H 1996  \textit{Plasma Phys. Control. Fusion} \href{http://iopscience.iop.org/article/10.1088/0741-3335/38/2/001/meta} {\textbf{38} 105}
\bibitem{ref7} Zhao N \emph{et al} 2018 \textit{Plasma Sci. Technol.} \href{http://iopscience.iop.org/article/10.1088/2058-6272/aa9bd8/meta} {\textbf{20} 024007}
\bibitem{ref8} Evans T E \emph{et al} 2006 \textit{Nat. Phys.} \href{http://dx.doi.org/10.1038/nphys312} {\textbf{2} 419}
\bibitem{ref9} Cai H S \emph{et al} 2018 \textit{Nucl. Fusion} \href{https://doi.org/10.1088/1741-4326/aaa55e} {\textbf{58} 036008}
\bibitem{ref10} Wei L and Wang Z X 2014 \textit{Nucl. Fusion} \href{https://doi.org/10.1088/0029-5515/54/4/043015} {\textbf{54} 043015}
\bibitem{ref11} Yang S X \emph{et al} 2018 \textit{Nucl. Fusion} \href{http://stacks.iop.org/0029-5515/58/i=4/a=046016} {\textbf{58} 046016}
\bibitem{ref12} Han M K \emph{et al} 2017 \textit{Nucl. Fusion} \href{https://doi.org/10.1088/1741-4326/aa5d02} {\textbf{57} 046019}
\bibitem{ref13} Chen W \emph{et al} 2010 \textit{Nucl. Fusion} \href{https://doi.org/10.1088/0029-5515/50/8/084008} {\textbf{50} 084008}
\bibitem{ref14} Kikuchi M and Azumi M 2012 \textit{Rev. Mod. Phys.} \href{http://dx.doi.org/10.1103/RevModPhys.84.1807} {\textbf{84} 1807}
\bibitem{ref15} La Haye R J \emph{et al} 2000 \textit{Phys. Plasmas} \href{http://dx.doi.org/10.1063/1.874199} {\textbf{7} 3349}
\bibitem{ref16} Gao X \emph{et al} 2015 \textit{Plasma Sci. Technol.} \href{http://stacks.iop.org/1009-0630/17/i=6/a=448} {\textbf{17} 448}
\bibitem{ref17} Gong X Z \emph{et al} 2017 \textit{Plasma Sci. Technol.} \href{http://stacks.iop.org/1009-0630/19/i=3/a=032001} {\textbf{19} 032001}
\bibitem{ref18} Sun Y W \emph{et al} 2016 \textit{Phys. Rev. Lett.} \href{http://dx.doi.org/10.1103/PhysRevLett.117.115001} {\textbf{117} 115001}
\bibitem{ref19} Nishimura S \emph{et al} 2010 \textit{Plasma Fusion Res.} \href{https://doi.org/10.1585/pfr.5.040} {\textbf{5} 040}
\bibitem{ref20} Hu Q M and Yu Q Q 2016 \textit{Nucl. Fusion} \href{http://dx.doi.org/10.1088/0029-5515/56/3/034001} {\textbf{56} 034001}
\bibitem{ref21} Wang J L, Wang Z X and Wei L 2015 \textit{Phys. Plasmas} \href{https://doi.org/10.1063/1.4931067} {\textbf{22} 092122}
\bibitem{ref22} Volpe F A \emph{et al} 2015 \textit{Phys. Rev. Lett.} \href{https://doi.org/10.1103/PhysRevLett.115.175002} {\textbf{115} 175002}
\bibitem{ref23} Hender T C \emph{et al} 1992 \textit{Nucl. Fusion} \href{http://dx.doi.org/10.1088/0029-5515/32/12/I02} {\textbf{32} 2091}
\bibitem{ref24} Yu Q Q, G$\rm \ddot{u}$nter S and Lackner K 2000 \textit{Phys. Rev. Lett.} \href{https://doi.org/10.1103/PhysRevLett.85.2949} {\textbf{85} 2949}
\bibitem{ref25} Hu Q M \emph{et al} 2012 \textit{Nucl. Fusion} \href{http://dx.doi.org/10.1088/0029-5515/52/8/083011} {\textbf{52} 083011}
\bibitem{ref26} Hu Q M \emph{et al} 2013 \textit{Phys. Plasmas} \href{http://dx.doi.org/10.1063/1.4820800} {\textbf{20} 092502}
\bibitem{ref27} Yu Q Q, G$\rm \ddot{u}$nter S and Lackner K 2018 \textit{Nucl. Fusion} \href{https://doi.org/10.1088/1741-4326/aab2fb} {\textbf{58} 054003}
\bibitem{ref28} Wang X G \emph{et al} 2015 \textit{Nucl. Fusion} \href{http://dx.doi.org/10.1088/0029-5515/55/9/093024} {\textbf{55} 093024}
\bibitem{ref29} Choi W \emph{et al} 2018 \textit{Nucl. Fusion} \href{https://doi.org/10.1088/1741-4326/aaa6e3} {\textbf{58} 036022}
\bibitem{ref30} Nave M F F and Wesson J A 1990 \textit{Nucl. Fusion} \href{http://dx.doi.org/10.1088/0029-5515/30/12/011} {\textbf{30} 2575}
\bibitem{ref31} Yu Q Q and G$\rm \ddot{u}$nter S \emph{et al} 2008 \textit{Nucl. Fusion} \href{http://dx.doi.org/10.1088/0029-5515/48/6/065004} {\textbf{48} 065004}
\bibitem{ref32} Yu Q Q \emph{et al} 2008 \textit{Nucl. Fusion} \href{http://dx.doi.org/10.1088/0029-5515/48/2/024007} {\textbf{48} 024007}
\bibitem{ref33} Sun Y W \emph{et al} 2010 \textit{Plasma Phys. Control. Fusion} \href{http://dx.doi.org/10.1088/0741-3335/52/10/105007} {\textbf{52} 105007}
\bibitem{ref34} Fitzpatrick R 2015 \textit{Phys. Plasmas} \href{http://dx.doi.org/10.1063/1.4919030} {\textbf{22} 042514}
\bibitem{ref35} Fitzpatrick R 2003 \textit{Phys. Plasmas} \href{http://dx.doi.org/10.1063/1.1560924} {\textbf{10} 1782}
\bibitem{ref36} Fitzpatrick R 1998 \textit{Phys. Plasmas} \href{https://doi.org/10.1063/1.873000 } {\textbf{5} 3325}
\bibitem{ref37} Hazeltine R D, Kotschenreuther M and Morrison P J 1985\textit{Phys. Fluids} \href{https://doi.org/10.1063/1.865255} {\textbf{28} 2466}
\bibitem{ref38} Strauss H R 1976 \textit{Phys. Fluids} \href{https://doi.org/10.1063/1.861310} {\textbf{19} 134}
\bibitem{ref39} Yu Q Q and G$\rm \ddot{u}$nter S 1998 \textit{Phys. Plasmas} \href{http://dx.doi.org/10.1063/1.873112} {\textbf{5} 3924}
\bibitem{ref40} Liu T \emph{et al} 2018 \textit{Nucl. Fusion} \href{https://doi.org/10.1088/1741-4326/aac527} {\textbf{58} 076026}
\bibitem{ref41} Ishii Y, Azumi M and Kishimoto Y 2002 \textit{Phys. Rev. Lett.} \href{http://dx.doi.org/10.1103/PhysRevLett.89.205002} {\textbf{89} 205002}
\bibitem{ref42} Bierwage A \emph{et al} 2005 \textit{Phys. Rev. Lett.} \href{http://dx.doi.org/10.1103/PhysRevLett.94.065001} {\textbf{94} 065001}
\bibitem{ref43} Wang Z X, Wei L and Yu F \emph{et al} 2015 \textit{Nucl. Fusion} \href{http://dx.doi.org/10.1088/0029-5515/55/4/043005} {\textbf{55} 043005}
\bibitem{ref44} Wei L \emph{et al} 2016 \textit{Nucl. Fusion} \href{https://doi.org/10.1088/0029-5515/56/10/106015} {\textbf{56} 106015}
\bibitem{ref45} Wang J L \emph{et al} 2017 \textit{Nucl. Fusion} \href{https://doi.org/10.1088/1741-4326/aa598c} {\textbf{57} 046007}
\bibitem{ref46} Sato M and Wakatani M 2005 \textit{Nucl. Fusion} \href{https://doi.org/10.1088/0029-5515/45/2/008} {\textbf{45} 143}
\end{thebibliography}
\end{document}